\documentclass[showpacs,aps,prc,preprint,nofootinbib,superscriptaddress]{revtex4}

\usepackage[utf8]{inputenc}
\usepackage{graphicx}
\usepackage{float}
\usepackage{amsmath}
\usepackage{amssymb}
\usepackage{color}
\usepackage{slashed}

\def\bm{\boldsymbol}
\newcommand{\bea}{\begin{eqnarray}}
\newcommand{\eea}{\end{eqnarray}}
\newcommand{\be}{\begin{eqnarray}}
\newcommand{\ee}{\end{eqnarray}}
\newcommand{\no}{\nonumber \\}

\newcommand{\sll}[1]{#1\hspace{-0.5em}/}

\def\vp{{\bm p}}

\def\vx{{\bm x}}
\def\vy{{\bm y}}

\def\vr{{\bm r}}
\def\vs{{\bm\sigma}}
\def\la{\langle}
\def\ra{\rangle}

\begin{document}


\title{Nuclear electric dipole moment of three-body system}


\author{Young-Ho Song}
\email[]{song25@mailbox.sc.edu}
\affiliation{Department of Physics and Astronomy,
University of South Carolina, Columbia, South Carolina, 29208, USA}

\author{Rimantas Lazauskas}
\email[]{rimantas.lazauskas@ires.in2p3.fr}
\affiliation{IPHC, IN2P3-CNRS/Universit\'e Louis Pasteur BP 28,
F-67037 Strasbourg Cedex 2, France}

\author{Vladimir Gudkov}
\email[]{gudkov@sc.edu}
\affiliation{Department of Physics and Astronomy,
University of South Carolina, Columbia, South Carolina, 29208, USA}




\date{\today}

\begin{abstract}
 Nuclear electric dipole moments  of $^3\mbox{He}$ and $^3\mbox{H}$ are calculated using
Time Reversal Invariance Violating (TRIV) potentials based on the meson exchange theory,
as well as the ones derived by using pionless and pionful effective field theories,
with nuclear  wave functions  obtained by solving Faddeev equations in configuration space
for the complete Hamiltonians comprising both TRIV and  realistic strong interactions.
The obtained results are compared with the previous calculations of $^3\mbox{He}$ EDM and with
time reversal invariance violating effects in neutron-deuteron scattering.
\end{abstract}


\maketitle

\section{Introduction
\label{sec:Int}}

The electric dipole moment (EDM) of particles
is a very important parameter in
searching for Time Reversal Invariance Violation (TRIV)
and for the possible manifestation of new physics.
The discovery of non-zero value of the EDM would
be a clear evidence of TRIV  \cite{Landau:1957},
therefore, it has been  a subject for intense  experimental
and theoretical investigations  for more than 50 years.
The search for TRIV also has fundamental importance 
for the explanation of the baryon asymmetry~\cite{Beringer:1900zz}
of the Universe which  requires a source of CP-violation \cite{Sakharov:1967dj}
beyond that entering the Cabibbo-Kobayashi- Maskawa
matrix of the Standard Model. Any observation of EDM
in the near future will be a direct indication of
new physics beyond the Standard Model.
However,  theoretical estimates for the values of particle EDMs
are extremely small which 
results in  many difficulties in experimental search
for neutron and electron EDMs.
Therefore, it is desirable to consider more complex systems,
where EDMs or another TRIV parameters could be enhanced.
This also would provide
assurance that there would be enough observations to avoid a possible ``accidental'' cancelation
of T-violating effects due to unknown structural factors related to  strong interactions.
The study of the EDM is  particularly important for the simplest few-nucleon systems, the result of 
which may lead to a better understanding of TRIV effects in heavier nuclei. Moreover,
few-nucleon systems meet requirements for a number of proposals to measure EDMs of light nuclei in storage rings~\cite{Khriplovich:1998zq,Farley:2003wt,Semertzidis:2009zz,Lehrach:2012eg}.

 In this paper, we calculate nuclear EDMs of $^3\mbox{He}$ and $^3\mbox{H}$
using TRIV potential in meson exchange model, as well as in pionless and pionful EFT.
Weak TRIV potentials are used in conjunction with  realistic strong interaction Hamiltonians.
Several realistic nucleon-nucleon potentials have been tested to represent the strong interaction: the Argonne v18(AV18), the Reid soft core(Reid93), the
Nijmegen(NijmII), the INOY, as well as the AV18
 in conjunction with   the three-nucleon Urbanna IX(UIX) potential.
Three-nucleon wave functions have been obtained
by solving Faddeev equations in the configuration space
for the complete Hamiltonians, comprising
both TRIV and strong interactions.

\section{Time reversal violating  potentials
\label{sec:TVPVpot}}

The most general form of time reversal violating
and parity violating part of nucleon-nucleon Hamiltonian
in the first order of relative nucleon momentum
can be written as the sum of momentum independent  and momentum dependent parts,
$H^{\slashed{T}\slashed{P}}=H^{\slashed{T}\slashed{P}}_{stat}+H^{\slashed{T}\slashed{P}}_{non-static}$
\cite{Pherzeg66},
\bea
\label{eq:static:pot}
H^{\slashed{T}\slashed{P}}_{stat}&=&g_1(r)\vs_{-}\cdot\hat{r}
                 +g_2(r)\tau_1\cdot\tau_2 \vs_{-}\cdot\hat{r}
                 +g_3(r)T_{12}^z\vs_{-}\cdot\hat{r}\no & &
                 +g_4(r)\tau_{+}\vs_{-}\cdot\hat{r}
                 +g_5(r)\tau_{-}\vs_{+}\cdot\hat{r}
\eea
\bea
\label{eq:nonstatic:pot}
H^{\slashed{T}\slashed{P}}_{non-static}
 &=&\left( g_6(r)+g_7(r)\tau_1\cdot\tau_2+g_8(r)T_{12}^z+g_9(r)\tau_{+} \right)\vs_\times\cdot\frac{\bar{\vp}}{m_N}
\no
 & &+\left(g_{10}(r)+g_{11}(r)\tau_1\cdot\tau_2+g_{12}(r)T_{12}^z+g_{13}(r)\tau_{+}\right)
 \no& &\times \left(\hat{r}\cdot\vs_\times\hat{r}
     \cdot \frac{\bar\vp}{m_N}
           -\frac{1}{3}\vs_\times\cdot \frac{\bar\vp}{m_N}\right)
  \no
 & &+g_{14}(r)\tau_{-}\Big(
    \hat{r}\cdot\vs_1\hat{r}\cdot(\vs_2\times \frac{\bar\vp}{m_N})
  +\hat{r}\cdot\vs_2\hat{r}\cdot(\vs_1\times \frac{\bar\vp}{m_N})
 \Big)\no
 & &+g_{15}(r)(\tau_1\times\tau_2)^z\vs_{+}\cdot \frac{\bar\vp}{m_N}
 \no & &+g_{16}(r)(\tau_1\times\tau_2)^z
    \left(\hat{r}\cdot\vs_{+}\hat{r}\cdot \frac{\bar\vp}{m_N}
         -\frac{1}{3}\vs_{+}\cdot \frac{\bar\vp}{m_N}
    \right),
\eea
where the exact form of $g_i(r)$ depends on the details of the particular theory. Because of the additional factor,
the non-static potential contributions are suppressed by a factor $\frac{\bar\vp}{m_N}$, therefore,
 we consider here only static TRIV interactions
 which could be obtained within three different approaches:
in a meson exchange model, pionless EFT, and pionful EFT.

The TRIV meson exchange potential in general
  involves  exchanges of
pions ($J^P=0^-$, $m_\pi=138$ MeV), $\eta$-mesons($J^P=0^-$, $m_\eta=550$ MeV), and
$\rho$- and $\omega$-mesons ($J^P=1^-$, $m_{\rho,\omega}=770,780$ MeV).
 To derive this potential, one can use strong ${\cal L}^{st}$ and TRIV ${\cal L}_{\slashed{T}\slashed{P}}$ Lagrangians, which can be written as~\cite{Herczeg:1987gp,Liu:2004tq}
\bea
{\cal L}^{st}&=&g_{\pi}\bar{N} i\gamma_5\tau^a \pi^a N
               +g_{\eta }\bar{N}i\gamma_5\eta N
             \no & &  -g_{\rho }\bar{N}\left(\gamma^\mu-i\frac{\chi_V}{2 m_N}\sigma^{\mu\nu} q_\nu\right)\tau^a \rho^a_\mu N
             \no & &  -g_{\omega }\bar{N}\left(\gamma^\mu-i\frac{\chi_S}{2m_N}\sigma^{\mu\nu} q_\nu\right)\omega_\mu N,
\eea
 \bea
{\cal L}_{\slashed{T}\slashed{P}}
&=&\bar{N}[\bar{g}_\pi^{(0)} \tau^a \pi^a+\bar{g}_\pi^{(1)}\pi^0
           +\bar{g}_\pi^{(2)}(3\tau^z\pi^0-\tau^a\pi^a)]N\no
& &+\bar{N}[\bar{g}^{(0)}_\eta\eta+\bar{g}^{(1)}_\eta \tau^z \eta] N\no
& &+\bar{N}\frac{1}{2m_N}[\bar{g}_\rho^{(0)}\tau^a \rho_\mu^a
                         +\bar{g}^{(1)}_\rho \rho^0_\mu
                         +\bar{g}^{(2)}(3\tau^z\rho_\mu^0-\tau^a\rho^a_\mu )]
                         \sigma^{\mu\nu}q_\nu\gamma_5 N \no
& &+\bar{N}\frac{1}{2m_N}[\bar{g}^{(0)}_\omega\omega_\mu
                         +\bar{g}^{(1)}_\omega \tau^z \omega_\mu]
                         \sigma^{\mu\nu}q_\nu\gamma_5 N,
\eea
where $q_\nu=p_\nu-p'_\nu$,
$\chi_V$ and $\chi_S$ are iso-vector and
scalar magnetic moments of a nucleon ($\chi_V=3.70$ and $\chi_S=-0.12$), and $\bar{g}^{(i)}_\alpha$ are TRIV meson-nucleon coupling constants.

Then, a TRIV potential obtained from these Lagrangians can be written as
\bea
\label{eq:pot}
V_{\slashed{T}\slashed{P}}
&=&
\left[-\frac{\bar{g}^{(0)}_\eta g_\eta}{2 m_N}
               \frac{m_\eta^2}{4\pi} Y_{1}(x_\eta)
               +\frac{\bar{g}^{(0)}_\omega g_\omega}{2 m_N}
               \frac{m_\omega^2}{4\pi}Y_{1}(x_\omega)\right]
               \vs_{-}\cdot\hat{r}
               \no
& &+\left[-\frac{\bar{g}^{(0)}_\pi g_\pi}{2 m_N}
              \frac{m_\pi^2}{4\pi} Y_{1}(x_\pi)
              +\frac{\bar{g}^{(0)}_\rho g_\rho}{2 m_N}
              \frac{m_\rho^2}{4\pi}Y_{1}(x_\rho)\right]
              \tau_1\cdot\tau_2\vs_{-}\cdot\hat{r}
              \no
& &+\left[-\frac{\bar{g}^{(2)}_\pi g_\pi}{2 m_N}
              \frac{m_\pi^2}{4\pi}Y_{1}(x_\pi)
              +\frac{\bar{g}^{(2)}_\rho g_\rho}{2 m_N}
              \frac{m_\rho^2}{4\pi}Y_{1}(x_\rho)\right]
              T_{12}^z\vs_{-}\cdot\hat{r}
              \no
& &+\left[-\frac{\bar{g}^{(1)}_\pi g_\pi}{4 m_N}
               \frac{m_\pi^2}{4\pi} Y_{1}(x_\pi)
              +\frac{\bar{g}^{(1)}_\eta g_\eta}{4 m_N}
               \frac{m_\eta^2}{4\pi} Y_{1}(x_\eta)
              +\frac{\bar{g}^{(1)}_\rho g_\rho}{4 m_N}
                \frac{m_\rho^2}{4\pi}Y_{1}(x_\rho)
              +\frac{\bar{g}^{(1)}_\omega g_\omega}{4 m_N}
               \frac{m_\omega^2}{4\pi} Y_{1}(x_\omega)\right]
               \tau_{+}\vs_{-}\cdot\hat{r}
              \no
& &+\left[-\frac{\bar{g}^{(1)}_\pi g_\pi}{4 m_N}
              \frac{m_\pi^2}{4\pi} Y_{1}(x_\pi)
              -\frac{\bar{g}^{(1)}_\eta g_\eta}{4 m_N}
              \frac{m_\eta^2}{4\pi} Y_{1}(x_\eta)
              -\frac{\bar{g}^{(1)}_\rho g_\rho}{4 m_N}
              \frac{m_\rho^2}{4\pi}Y_{1}(x_\rho)
              +\frac{\bar{g}^{(1)}_\omega g_\omega}{4 m_N}
              \frac{m_\omega^2}{4\pi}Y_{1}(x_\omega)\right]
              \tau_{-}\vs_{+}\cdot\hat{r},\no
\eea
where $T_{12}^z=3\tau_1^z\tau_2^z-\tau_1\cdot\tau_2$,
$Y_1(x)=(1+\frac{1}{x})\frac{e^-x}{x}
=-\frac{d}{dx}Y_0(x)$,  $Y_0(x)=\frac{e^-x}{x}$
,$x_a=m_a r$.

Comparing  eq. (\ref{eq:static:pot}) with this potential, one can see that $g_i(r)$ functions in a 
meson exchange model can be identified as
\bea
\label{eq:gi:ME}
g_1^{ME}(r)&=&-\frac{\bar{g}^{(0)}_\eta g_\eta}{2 m_N}
               \frac{m_\eta^2}{4\pi} Y_{1}(x_\eta)
               +\frac{\bar{g}^{(0)}_\omega g_\omega}{2 m_N}
               \frac{m_\omega^2}{4\pi}Y_{1}(x_\omega)\no
g_2^{ME}(r)&=&-\frac{\bar{g}^{(0)}_\pi g_\pi}{2 m_N}
              \frac{m_\pi^2}{4\pi} Y_{1}(x_\pi)
              +\frac{\bar{g}^{(0)}_\rho g_\rho}{2 m_N}
              \frac{m_\rho^2}{4\pi}Y_{1}(x_\rho)
              \no
g_3^{ME}(r)&=&-\frac{\bar{g}^{(2)}_\pi g_\pi}{2 m_N}
              \frac{m_\pi^2}{4\pi}Y_{1}(x_\pi)
              +\frac{\bar{g}^{(2)}_\rho g_\rho}{2 m_N}
              \frac{m_\rho^2}{4\pi}Y_{1}(x_\rho)
              \no
g_4^{ME}(r)&=&-\frac{\bar{g}^{(1)}_\pi g_\pi}{4 m_N}
               \frac{m_\pi^2}{4\pi} Y_{1}(x_\pi)
              +\frac{\bar{g}^{(1)}_\eta g_\eta}{4 m_N}
               \frac{m_\eta^2}{4\pi} Y_{1}(x_\eta)
              +\frac{\bar{g}^{(1)}_\rho g_\rho}{4 m_N}
                \frac{m_\rho^2}{4\pi}Y_{1}(x_\rho)
              +\frac{\bar{g}^{(1)}_\omega g_\omega}{4 m_N}
               \frac{m_\omega^2}{4\pi} Y_{1}(x_\omega)
              \no
g_5^{ME}(r)&=&-\frac{\bar{g}^{(1)}_\pi g_\pi}{4 m_N}
              \frac{m_\pi^2}{4\pi} Y_{1}(x_\pi)
              -\frac{\bar{g}^{(1)}_\eta g_\eta}{4 m_N}
              \frac{m_\eta^2}{4\pi} Y_{1}(x_\eta)
              -\frac{\bar{g}^{(1)}_\rho g_\rho}{4 m_N}
              \frac{m_\rho^2}{4\pi}Y_{1}(x_\rho)
              +\frac{\bar{g}^{(1)}_\omega g_\omega}{4 m_N}
              \frac{m_\omega^2}{4\pi}Y_{1}(x_\omega),\no
\eea

The TRIV potentials in pionless EFT (without explicit pion contributions) contain only point-like nucleon-nucleon interactions which are proportional to the local
delta functions. Thus, one can write the corresponding $g_i(r)$ functions as
\bea
\label{eq:gi:pionless}
g_{i=1..5}^{\sll{\pi}}(r)
=\frac{c_{i=1..5}^{\sll{\pi}}}{2m_N}\frac{d}{dr}\delta^{(3)}(r)
\to \frac{c_{i=1..5}^{\sll{\pi}}\mu^2}{2m_N}
    \left(-\frac{\mu^2}{4\pi}Y_1(\mu r)\right),
\eea
where low energy constants (LECs) $c_i^{\sll{\pi}}$
of pionless EFT have the dimension of $[fm^2]$.
Here, for numerical calculations we approximate the singular delta functions by the Yukawa type functions
as $\delta^{(3)}(r)\simeq\frac{\mu^3}{4\pi}Y_0(\mu r)$, as it was done in~\cite{Liu:2004tq},
with a natural scale of the parameter  $\mu\simeq m_\pi$.

In the pionful EFT, the long range terms of the potential are due to the one pion exchange whereas
the short range terms are similar to the ones obtained within the pionless EFT. Then, by ignoring the contribution of the two pion exchange
at the middle range scale, as well as other higher order corrections, one can write $g_i(r)$ functions for the pionful EFT as
\bea
\label{eq:gi:pionful}
g_1^{\pi}(r)&=& -\frac{c_1^{\pi} \mu^2}{2 m_N} \frac{\mu^2}{4\pi}Y_1(\mu r)\no
g_2^{\pi}(r)&=& -\frac{c_2^{\pi} \mu^2}{2 m_N} \frac{\mu^2}{4\pi}Y_1(\mu r)-\frac{\bar{g}^{(0)}_\pi g_\pi}{2 m_N}
              \frac{m_\pi^2}{4\pi} Y_{1}(x_\pi)\no
g_3^{\pi}(r)&=& -\frac{c_3^{\pi} \mu^2}{2 m_N} \frac{\mu^2}{4\pi}Y_1(\mu r)-\frac{\bar{g}^{(2)}_\pi g_\pi}{2 m_N}
              \frac{m_\pi^2}{4\pi}Y_{1}(x_\pi)\no
g_4^{\pi}(r)&=& -\frac{c_4^{\pi} \mu^2}{2 m_N} \frac{\mu^2}{4\pi}Y_1(\mu r)-\frac{\bar{g}^{(1)}_\pi g_\pi}{4 m_N}
               \frac{m_\pi^2}{4\pi} Y_{1}(x_\pi)\no
g_5^{\pi}(r)&=& -\frac{c_5^{\pi} \mu^2}{2 m_N} \frac{\mu^2}{4\pi}Y_1(\mu r)
-\frac{\bar{g}^{(1)}_\pi g_\pi}{4 m_N}
              \frac{m_\pi^2}{4\pi} Y_{1}(x_\pi).
\eea
For this potential, the cutoff scale $\mu$ is larger than pion mass, because pion is an explicit degree of freedom of the theory.\footnote{These expressions  do not include the cutoff for the pion exchange terms.
The introduction of this cutoff with the corresponding Fourier
transformation  will modify the Yukawa functions.  This will modify the
contributions from the short distance terms, as well as the form of
the scalar functions in contact terms. However, these corrections are of
a higher order.
It should be mentioned that
the values of the LECs and their scaling behavior,
as a function of a cutoff parameter
$c_i^{\pi}(\mu)$, differ from
the corresponding behavior of $c_i^{\not\pi}(\mu)$ terms obtained in pionless EFT.}
The following identities might be useful in order to compare this potential with the one used in reference~\cite{deVries:2011an}:
\bea
\frac{g_\pi}{2m_N}\leftrightarrow \frac{g_A}{F_\pi}
\mbox{ in \cite{deVries:2011an} } ,\quad
\bar{g}^{(0,1)}_\pi
\leftrightarrow
\left(-\frac{\bar{g}_{0,1}}{F_\pi}\right)
\mbox{ in \cite{deVries:2011an} }, \quad
\frac{c_{1,2}^\pi}{2m_N}\leftrightarrow
\frac{1}{2}\bar{C}_{1,2} \mbox{ in \cite{deVries:2011an} }.
\eea
However, the parameters  $c_{3,4,5}^\pi$ and $\bar{g}^{(2)}_\pi$
were not included at the leading order potential
in~\cite{deVries:2011an} because
they were considered as higher order terms
with  additional assumptions related to the source of TRIV interactions.
 In this paper,  we calculate contributions from all the operators without
making any assumption about the possible value of the coefficients for  each operator.

It is important  that  all these three potentials which come from different approaches have exactly the same operator structure.
Thus, the only difference between them  is related to the difference in corresponding scalar  functions
which, in turn, differ only by the values of the characteristic masses: $m_\pi$, $m_\eta$, $m_\rho$, and $m_\omega$.
Therefore, to unify the notation, it is convenient to define the new  constants $C_n^a$ (having dimension of $[fm]$)
together with the scalar function $f_n^a(r)=\frac{\mu^2}{4\pi}Y_1(\mu r)$ (having dimension of $[fm^{-2}]$) as
\bea
g_n(r)\equiv \sum_{a} C_n^a f_n^a(r),
\eea
where the expressions of $C_n^a$ and $f_n^a(r)$ are provided in eqs.
(\ref{eq:gi:ME}), (\ref{eq:gi:pionless}), and (\ref{eq:gi:pionful}).

Since the non-static TRIV potential terms, with $g_{n>5}$,
do not appear  either in a meson exchange model
or in the lowest order EFTs,
they can be considered as a higher order correction
to the lowest order EFT
or be related to heavy  meson or
multi-meson contributions in the meson exchange model.

\section{$^3\mbox{He}$ and $^3\mbox{H}$ EDMs}
The value of nuclear EDM is defined as
\bea
d=\la JJ|\hat{D}|JJ\ra
 =\sqrt{\frac{J}{(2J+1)(J+1)}}\la J||\hat{D}||J\ra ,
\eea
where $|JJ\ra$ is a nuclear wave function with a total spin and its projection equal to $J$.
The EDM operator $\hat{D}$ contains direct contributions
from the intrinsic nucleon EDMs (current operators)
\bea
\label{eq:intr_edm}
\hat{D}^{nucleon}_{\sll{T}\sll{P}}
=\sum_i \frac{1}{2}[(d_p+d_n)+(d_p-d_n)\tau_i^z]\vs_i
\eea
and contributions from the nuclear EDM polarization operator
\bea
\hat{D}^{pol}_{PT}
=\sum_i Q_i \vr_i ,
\eea
which describe polarization of the nuclei due to TRIV potentials.
Here, $d_n$ and $d_p$ are neutron and proton EDMs, and $Q_i$ and $r_i$ are charge and position of $i$-th nucleon.
In this work, we do not consider possible TRIV
meson exchange current contributions.
Therefore, the value of $^3\mbox{He}$ (or $^3\mbox{H}$) EDM can  be expressed as
\bea
d=\frac{1}{\sqrt{6}}
  \left[\la \Psi||\hat{D}^{nucleon}_{\sll{T}\sll{P}}||\Psi\ra
       +\la \Psi_{\sll{T}\sll{P}}||\hat{D}^{pol}_{TP}||\Psi\ra
       +\la \Psi||\hat{D}^{pol}_{TP}||\Psi_{\sll{T}\sll{P}}\ra
  \right],
\eea
where $|\Psi\ra$ and $|\Psi_{\sll{T}\sll{P}}\ra$ represent the time reversal invariant and TRIV parts of the nuclear wave function.

 First, we will analyze  the contribution of the intrinsic nucleon EDM momenta to
the nuclear EDM which are defined by the action of operator defined
in eq.(\ref{eq:intr_edm}) on the nuclear wave function.
For  the deuteron  (the two-body system),  the contributions of the intrinsic neutron and proton EDMs simply add
$d^{nucleon}_{d}=d_p+d_n$, i.e. this value does not depend on the nuclear wave function and, consequently,  on the choice of the particular strong interaction model.
The situation is  different in the three-body system, $^3\mbox{H}$ or $^3\mbox{He}$ nuclei, where
the EDM contributions from intrinsic nucleon EDMs become wave function dependent.
One can see that nucleonic contributions to the nuclear EDMs are in rather good agreement for all strong potentials with local interactions: the AV18, the Reid93 and the Nijm II.
Nevertheless, these values  differ for the models that includ non-locality or for the ones where a three-nucleon force is added (see Table \ref{tbl:nucleons}
where EDM calculations from the references~\cite{Stetcu:2008vt,deVries:2011an} are also listed).
Softer two-nucleon interaction models\footnote{The INOY interaction model has the strongest non-locality and is the softest one from the interactions considered in the table~\ref{tbl:nucleons},  followed by the EFT and the CD-BONN potentials, respectively.} have a tendency to provide the nuclear EDM values closer to the ones of the unpaired nucleon (i.e. neutronic EDM for $^3\mbox{He}$ case, and protonic EDM for $^3\mbox{H}$). The addition of the three-nucleon force provides an effect similar to the one of hardening the interaction. This effect is clearly related to the strength of the tensor force, which permits one to reduce the pairing of the nucleons in the three-nucleon system.
Let us mention that our results for this single-nucleonic operator are in excellent agreement with those from  references~\cite{Stetcu:2008vt,deVries:2011an}.

\begin{table}[H]
\caption{\label{tbl:nucleons}The nucleon electric
dipole moment contributions to nuclear EDMs calculated for different strong interaction potentials.
}
\begin{center}
\begin{tabular}{c|c|c}
model & $^3\mbox{He}$ & $^3\mbox{H}$ \\
      \hline
AV18    &$ -0.0468 d_p+0.877 d_n$ &$0.877 d_p-0.0480 d_n$  \\
Reid93  &$ -0.0465 d_p+0.878 d_n$ &$0.879 d_p-0.0475 d_n$ \\
NijmII  &$ -0.0458 d_p+0.880 d_n$ &$0.880 d_p-0.0468 d_n$ \\
AV18UIX &$ -0.0542 d_p+0.868 d_n$ &$0.868 d_p-0.0552 d_n$ \\
INOY    &$-0.0229d_p+0.927 d_n $ &$0.928 d_p-0.0236 d_n$ \\
\hline
CD-BONN\cite{Stetcu:2008vt,deVries:2011an} & $-0.0370 d_p+0.897 d_n$ &  -    \\
AV18\cite{Stetcu:2008vt,deVries:2011an}    & $-0.0470 d_p+0.877 d_n$ &  -   \\
EFT NN\cite{Stetcu:2008vt,deVries:2011an}  & $-0.0310 d_p+0.905 d_n$ &  -    \\
EFT NN+NNN\cite{Stetcu:2008vt,deVries:2011an} & $-0.0350 d_p+0.901 d_n$ & -  \\
\end{tabular}
\end{center}
\end{table}

In order to calculate polarization contributions to the nuclear EDM, we solve Faddeev equations in a configuration space~\cite{Faddeev:1960su}
by including TRIV potentials. We consider
 neutrons and protons as  isospin-degenerate states of the same particle nucleon whose
mass is fixed to $\hbar ^{2}/m=41.471$ MeV$\cdot$fm.  By using the isospin formalism, the
three Faddeev equations become formally identical, which for pairwise interactions reads
\begin{equation}
\left( E-H_{0}-V_{ij}\right) \psi _{k}
=V_{ij}(\psi _{i}+\psi _{j}),
\label{EQ_FE}
\end{equation}
where $(ijk)$ are particle indexes, $H_{0}$ is the kinetic energy operator,
$V_{ij}$ is a two body force between particles $i$, and $j$, and
$\psi _{k}=\psi_{ij,k}$ is the so-called Faddeev component.
In the last equation, the potential formally contains both a strong
interaction (TRI conserving) part ($V^{TC}_{ij}$) and a TRIV (parity violating)
 part ($V^{\sll{T}\sll{P}}$), i.e.: $V_{ij}=V^{TC}_{ij}+V^{{\sll{T}\sll{P}}}_{ij}$.
Due to the presence of
TRIV  potential, the system's wave function does not have a definite parity and contains both
 positive and negative parity components.
As a consequence, the Faddeev components of the total wave function can be split into the sum of positive- and negative-parity parts:
\begin{equation}
\psi _{k}=\psi^{+}_{k}+\psi^{-}_{k} .
\end{equation}
Three-nucleon bound state wave function has a strongly predominant positive-parity component.
The TRIV interaction is weak ($V^{{\sll{T}\sll{P}}}_{ij}<<V^{TC}_{ij}$). Then, by neglecting second-order terms in TRIV potential,  one obtains a system of two differential equations:
\begin{eqnarray}
\left( E-H_{0}-V^{TC}_{ij}\right) \psi^{+} _{k}
&=&V^{TC}_{ij}(\psi^{+} _{i}+\psi^{+} _{j}) \label{EQ_FE_1}, \\
\left( E-H_{0}-V^{TC}_{ij}\right) \psi^{-} _{k}
&=&V^{TC}_{ij}(\psi^{-} _{i}+\psi^{-} _{j})+V^{{\sll{T}\sll{P}}}_{ij}(\psi^{+} _{i}+\psi^{+} _{j}+\psi^{+} _{k}) \label{EQ_FE_2}
\end{eqnarray}
One can see that the
first equation (\ref{EQ_FE_1}) defines only the positive-parity part of the wave function.
This equation contains only a strong
nuclear potential and corresponds to the standard three-nucleon problem: a bound state of helium or triton.
The solution of the second differential equation
(\ref{EQ_FE_2}), which contains an inhomogeneous term $V^{{\sll{T}\sll{P}}}_{ij}(\psi^{+} _{i}+\psi^{+} _{j}+\psi^{+} _{k})$, gives us the negative-parity components of the wave functions.

To solve these equations numerically, we use our standard procedure
described in  detail in~\cite{These_Rimas_03}.
Using a set of Jacobi coordinates, defined by
$\vx_{k}=(\vr_{j}-\vr_{i})\smallskip $
and
$\vy_{k}=
\frac{2}{\sqrt{3}}(\vr_{k}-\frac{\vr_{i}+\vr_{j}}{2})$, we expand each Faddeev component of the wave function
 in bipolar harmonic basis:
\begin{equation}
\psi _{k}^{\pm}=\sum\limits_{\alpha }\frac{F^{\pm}_{\alpha }(x_{k},y_{k})}{x_{k}y_{k}}%
\left\vert \left( l_{x}\left( s_{i}s_{j}\right) _{s_{x}}\right)
_{j_{x}}\left( l_{y}s_{k}\right) _{j_{y}}\right\rangle _{JM}\otimes
\left\vert \left( t_{i}t_{j}\right) _{t_{x}}t_{k}\right\rangle _{TT_{z}},
\label{EQ_FA_exp}
\end{equation}%
where index $\alpha $ represents all allowed combinations of the
quantum numbers presented in the brackets, $l_{x}$ and $l_{y}$ are the
partial angular momenta associated with respective Jacobi coordinates, and
$s_{i} $ and $t_{i}$ are  spins and isospins of the individual particles. Functions
$F_{\alpha }(x_{k},y_{k})$ are called partial Faddeev amplitudes. In the expansion (\ref{EQ_FA_exp}), we
consider both possible total isospin channels $T=1/2$ and $T=3/2$, regardless of the
fact that positive-parity components $\psi _{k}^{+}$ have predominant contribution of  $T=1/2$ state.

Equations (\ref{EQ_FE_1}) and (\ref{EQ_FE_2}) must be supplemented  the appropriate boundary conditions for
Faddeev partial amplitudes $F_{\alpha }^{\pm}$:
 partial Faddeev amplitudes are regular at the origin
\begin{equation}
F_{\alpha }^{\pm}(0,y_{k})=F^{\pm}_{\alpha }(x_{k},0)=0,  \label{BC_xyz_0}
\end{equation}%
and the system's wave function  vanishes exponentially as either $x_{k}$
or $y_{k}$ becomes large. This condition is imposed by setting Faddeev amplitudes to vanish
at the borders $(x_{max},y_{max})$ of  a chosen grid, i.e.:
\begin{equation}
F_{\alpha }^{\pm}(x_{k},y_{max})=0,\quad
F^{\pm}_{\alpha }(x_{max},y_{k})=0.  \label{BC_xyz_0}
\end{equation}%
This formalism  can be easily generalized to accommodate three-nucleon forces, as is
described in paper~\cite{Lazauskas_2008}.

In Table \ref{tbl:pol}, we summarize the calculation of the matrix elements $\frac{2}{\sqrt{6}}
        \la \Psi||\hat{D}^{pol}_{TP}||\Psi_{\sll{T}\sll{P}}\ra$;
values for each TRIV operator from Eq.(\ref{eq:pot}) obtained for a different choice of the strong interaction are tabulated.
In this table, operators of TRIV potential (\ref{eq:pot}) are calculated when combined with a unified
scalar function
\bea
-\frac{1}{2m_N}\frac{\Lambda^2}{4\pi}Y_1(\Lambda r)
\eea
and calculated when taking various values of the parameter $\Lambda$, which was chosen to coincide
with the masses of $\pi$, $\eta$, $\rho$ and $\omega$ mesons.
Therefore, this table can be used to analyze the TRIV potentials in the meson exchange model, whereas once multiplied by an additional
$\Lambda^2$ cutoff factor, it also can be used for the TRIV potentials in pionless EFT or pionful EFT.

\begin{table}[H]
\caption{\label{tbl:pol} Contribution of the different TRIV operators (Eq.(\ref{eq:pot})
to the expectation value of
        $\frac{2}{\sqrt{6}}
        \la \Psi||\hat{D}^{pol}_{TP}||\Psi_{\sll{T}\sll{P}}\ra$. Calculations has been performed for
        several different strong potentials and for
          $^3\mbox{He}$ ($^3\mbox{He}$) nucleus; values are given
         in $10^{-3}$  e-fm units.
}
\begin{tabular}{ccccccc}
operator & $\Lambda$      & AV18      & Reid93  &NijmII & AV18UIX & INOY \\
\hline
$1$    & $m_\pi$   & $ -5.32( 5.28)$ & $ -5.37( 5.33) $& $ -5.31( 5.28) $& $ -4.46( 4.42)$ & $ -7.24( 7.23)$\\
         & $m_\eta$  & $-0.571(0.572)$ & $-0.608(0.609) $& $-0.584(0.585) $& $-0.478(0.477)$ & $ -1.53( 1.54)$\\
         & $m_\rho$  & $-0.233(0.234)$ & $ -0.26(0.261) $& $-0.241(0.242) $& $-0.195(0.195)$ & $-0.857(0.862)$\\
         & $m_\omega$& $-0.223(0.224)$ & $-0.249( 0.25) $& $-0.231(0.232) $& $-0.187(0.186)$ & $-0.833(0.838)$\\
\hline
$2$    & $m_\pi$   & $5.9  (-5.89  )$&$ 6.08 ( -6.07 )$&$ 6.12  ( -6.11 )$&$  5.5  ( -5.48 )$&$10.3 (-10.2 )$\\
         & $m_\eta$  & $0.673( -0.681)$&$ 0.803(  -0.81)$&$ 0.771 ( -0.777)$&$ 0.629 ( -0.635)$&$ 2.72( -2.73)$\\
         & $m_\rho$  & $0.292( -0.296)$&$ 0.387( -0.391)$&$ 0.351 ( -0.354)$&$  0.27 ( -0.273)$&$  1.6(  -1.6)$\\
         & $m_\omega$& $0.281( -0.284)$&$ 0.374( -0.378)$&$ 0.337 ( -0.341)$&$ 0.259 ( -0.262)$&$ 1.56( -1.56)$\\
\hline
$3$    & $m_\pi$   &$ 6.76( -7.02)$&$ 6.78( -7.01)$&$ 6.76( -6.98)$&$ 6.66( -6.89)$&$ 7.46( -7.72)$\\
         & $m_\eta$  &$0.775(-0.814)$&$0.773(-0.804)$&$0.762(-0.794)$&$0.784(-0.819)$&$ 1.25( -1.31)$\\
         & $m_\rho$  &$0.304( -0.32)$&$  0.3(-0.312)$&$0.295(-0.307)$&$0.308(-0.322)$&$0.645(-0.674)$\\
         & $m_\omega$&$ 0.29(-0.305)$&$0.285(-0.297)$&$0.281(-0.293)$&$0.294(-0.307)$&$0.625(-0.653)$\\
\hline
$4$    & $m_\pi$   &$2.17 (2.42 )$&$ 2.2 (2.41 )$&$2.25 (2.46 )$&$2.81 (3.03 )$&$2.27 (2.48 )$\\
         & $m_\eta$  &$0.286(0.319)$&$0.291(0.317)$&$0.296(0.322)$&$0.372(0.403)$&$0.397(0.436)$\\
         & $m_\rho$  &$0.112(0.125)$&$0.114(0.125)$&$0.116(0.127)$&$0.146(0.159)$&$0.202(0.223)$\\
         & $m_\omega$&$0.107( 0.12)$&$0.109(0.119)$&$0.111(0.121)$&$0.139(0.152)$&$0.196(0.216)$\\
\hline
$5$    & $m_\pi$   &  $  19.4( 19.6)$&$19.6(19.8)$&$  20(20.2)$&$ 18.3( 18.5)$&$19.5(19.6)$\\
         & $m_\eta$  &  $  2.43( 2.47)$&$2.59(2.63)$&$2.75( 2.8)$&$ 2.32( 2.35)$&$ 3.5(3.56)$\\
         & $m_\rho$  &  $ 0.985( 1.01)$&$1.09(1.11)$&$ 1.2(1.22)$&$0.937(0.953)$&$1.92(1.95)$\\
         & $m_\omega$&  $ 0.942(0.961)$&$1.04(1.06)$&$1.15(1.17)$&$0.896(0.911)$&$1.86( 1.9)$\\
\end{tabular}

\end{table}

\begin{figure}
\caption{\label{fig1}
 The relative deviations of the $d^{pol}_{^3\mbox{He}}$ value from the one obtained for AV18 potential $\frac{d^{pol}-d^{pol}(AV18)}{d^{pol}(AV18)}\times 100$. Results
 are presented for the operators 1(upper) and 5(lower) and as a function of the cutoff parameter.
}
\includegraphics[width=0.5\textwidth]{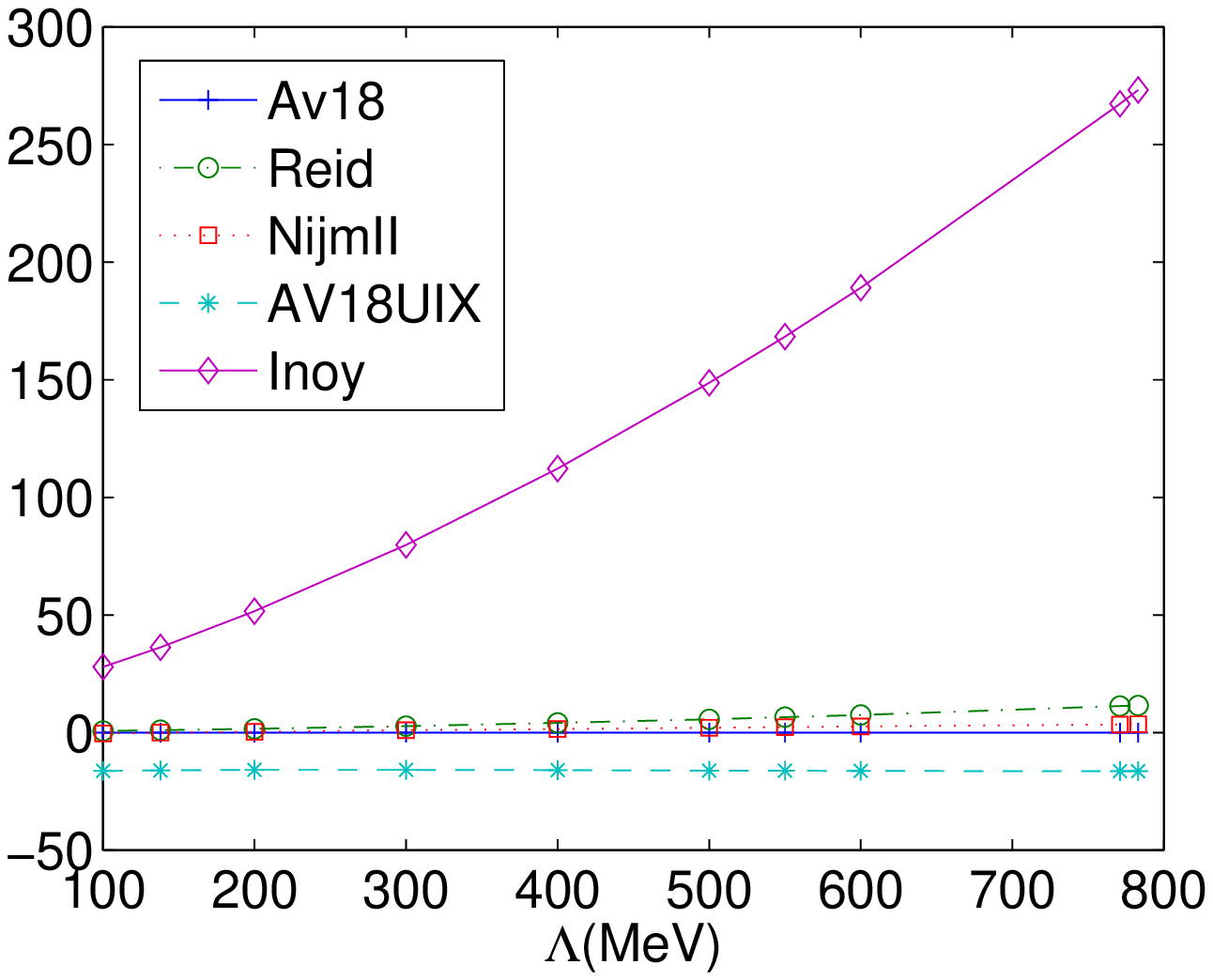}
\includegraphics[width=0.5\textwidth]{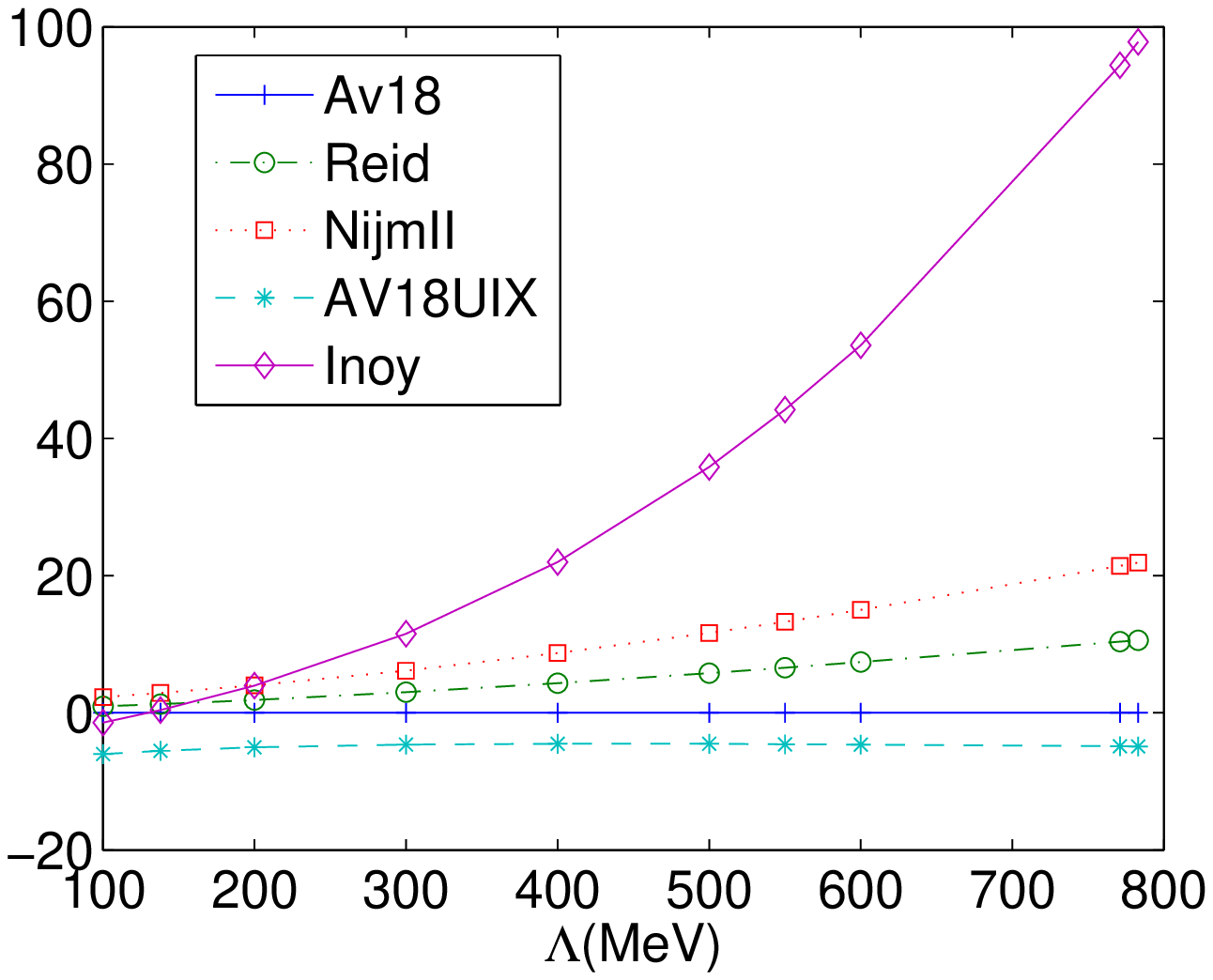}
\end{figure}

Similar to the intrinsic nucleon EDM contribution, presented in table~\ref{tbl:nucleons},  dynamical nuclear EDMs
are also rather insensitive to the choice of the local strong interaction potential. The addition of the three-nucleon
force affects the results by 10-20\%, whereas the presence of the strong non-locality in two-nucleon interaction (as in the INOY model) has a very large impact.
This fact is also confirmed by the cutoff dependence behavior of the relative deviations of the $\frac{2}{\sqrt{6}}
        \la \Psi||\hat{D}^{pol}_{TP}||\Psi_{\sll{T}\sll{P}}\ra $ matrix element for operators 1 and 5  calculated for different
         strong interaction potentials in relation to the matrix element calculated with AV18 potential (see  Fig. (\ref{fig1})),
          where the first plot corresponds to the operator 1 and second one to the operator 5.

Using meson exchange TRIV potential from Eq.(\ref{eq:pot}), we can present the results of the calculations of nuclear EDMs as the sum of contributions of the different TRIV potential terms to the
$\frac{2}{\sqrt{6}}\la \Psi||\hat{D}^{pol}_{TP}||\Psi_{\sll{T}\sll{P}}\ra$
matrix element, as  is shown in Table \ref{tbl:meson}.

\begin{table}[H]
\caption{\label{tbl:meson}Contributions to $\frac{2}{\sqrt{6}}
        \la \Psi||\hat{D}^{pol}_{TP}||\Psi_{\sll{T}\sll{P}}\ra$ for  $^3\mbox{He}(^3\mbox{H})$ EDMs from different terms of  meson exchange TRIV potential in $10^{-3}$ e-fm units.
We use the following values for strong couplings constants: $g_{\pi}=13.07$, $\quad g_{\eta}=2.24$, $\quad g_{\rho}=2.75$, $\quad g_{\omega}=8.25$. A similar table can be inferred for the case of pionless and pionful EFT from Table \ref{tbl:pol}.
}
\begin{tabular}{c|ccccc|c}
Couplings       & AV18   & Reid93 & NijmII &AV18UIX & INOY & AV18\cite{Stetcu:2008vt}\\
\hline
$\bar{g}_\pi^0$   &77.2(-76.9)  & 79.5(-79.3)  & 80.0(-79.8) &71.9(-71.6)  &134(-134)   & 157\\
$\bar{g}_\pi^1$   &141(144)     & 143(145)     & 145(148)    &138(141)     &142(145)    & 288\\
$\bar{g}_\pi^2$   &88.3(-91.8)  & 88.7(-91.6)  & 88.3(-91.3) &87.1(-90.1)  &98.5(-102)  & 444\\
$\bar{g}_\rho^0$  &-0.803(0.814)& -1.06(1.08)  &-0.964(0.974)&-0.742(0.751)&-4.40(4.41) &-1.65 \\
$\bar{g}_\rho^1$  & 1.20(1.21)  &  1.34(1.35)  & 1.49(1.50)  &1.09(1.09)   & 2.36(2.37) & 2.48\\
$\bar{g}_\rho^2$  &-0.836(0.879)& -0.824(0.858)&-0.811(0.845)&-0.846(0.885)&-1.77(1.85) & 4.13\\
$\bar{g}_\omega^0$& 1.84(-1.85) & 2.05(-2.06)  & 1.91(-1.91) &1.54(-1.54)  & 6.88(-6.91)& 4.13\\
$\bar{g}_\omega^1$&-4.33(-4.46) & -4.74(-4.86) &-5.19(-5.32) &-4.27(-4.38) &-8.49(-8.71)&-9.08\\
$\bar{g}_\eta^0$  &-1.28(1.28)  & -1.36(1.36)  &-1.31(1.31)  &-1.07(1.07)  &-3.43(3.45) & - \\
$\bar{g}_\eta^1$  & 2.40(2.41)  & 2.57(2.59)   & 2.75(2.77)  & 2.18(2.18)  & 3.48(3.50) & - \\
\end{tabular}
\end{table}

 The last column  in Table \ref{tbl:meson} shows the results for $^3\mbox{He}$ EDM obtained in reference~\cite{Stetcu:2008vt}.
Comparing results of~\cite{Stetcu:2008vt} with our calculations for AV18 potential, one can see that there is a systematic discrepancy for all the values of
the matrix elements.
For these calculations, we have used the same strong and TRIV potential as in~\cite{Stetcu:2008vt}. It, therefore, points at the possible systematic
error in one of the calculations. We use solutions of the Faddeev equations, while the calculations of the wave functions in~\cite{Stetcu:2008vt} have been done in a no-core shell model framework using perturbative expansion for the negative parity states. Based on the presented results,  it is impossible to figure out if the  discrepancy is the result of a numerical error in one of the algorithms or if there is an intrinsic limitation of the perturbative expansion used in reference~\cite{Stetcu:2008vt} with no-core shell model approach\footnote{Curiously enough, our results differ from the ones of reference~\cite{Stetcu:2008vt} roughly by the factor 2 for all the isospin rank-0 and rank-1 TRIV potential terms, whereas isospin rank-2 operator results differ by factor 5.}.

On the other hand, our calculations for the deuteron (two-body system) using the same formalism and employing
AV18 strong interaction gives
\bea
d_d^{(pol)}
= 18.95 \times 10^{-2}\bar{g}_\pi^1
  +3.52\times 10^{-3}\bar{g}_\eta^1
  +17.13\times 10^{-4}\bar{g}_\rho^1
  -49.09\times 10^{-4}\bar{g}_\omega^1,
\eea
which is in excellent agreement with the result of reference~\cite{Liu:2004tq}
\bea
d_d^{(pol)}
=18.69 \times 10^{-2} \bar{g}_\pi^1
+3.56\times 10^{-3} \bar{g}_\eta^1
+17.19\times 10^{-4} \bar{g}_\rho^1
-49.17\times 10^{-4} \bar{g}_\omega^1.
\eea

\section{Conclusions}

Using the data from Table \ref{tbl:meson}, one can obtain  the polarization parts of $^3\mbox{He}$ and $^3\mbox{H}$ EDMs for the choice of AV18UIX strong potential. Then, the expressions for $^3\mbox{He}$ and $^3\mbox{H}$ EDMs can be written as
\bea
\label{eq:3HeEDM}
d_{^3\mbox{He}}&=& (-0.0542 d_p+0.868 d_n)
+0.072 [ \bar{g}_\pi^{(0)}+1.92 \bar{g}_\pi^{(1)} +1.21 \bar{g}_\pi^{(2)}
-0.015 \bar{g}_\eta^{(0)}+0.03 \bar{g}_\eta^{(1)} \no & &
-0.010 \bar{g}_\rho^{(0)}+0.015 \bar{g}_\rho^{(1)}-0.012 \bar{g}_\rho^{(2)}
+0.021 \bar{g}_\omega^{(0)}-0.06 \bar{g}_\omega^{(1)}]{e\cdot fm}
\eea
and
\bea
\label{eq:3HEDM}
d_{^3\mbox{H}}&=& (0.868 d_p-0.0552 d_n)
-0.072
[ \bar{g}_\pi^{(0)}-1.97 \bar{g}_\pi^{(1)} +1.26 \bar{g}_\pi^{(2)}
-0.015 \bar{g}_\eta^{(0)}-0.030 \bar{g}_\eta^{(1)} \no & &
-0.010 \bar{g}_\rho^{(0)}-0.015 \bar{g}_\rho^{(1)}-0.012 \bar{g}_\rho^{(2)}
+0.022 \bar{g}_\omega^{(0)}+0.061 \bar{g}_\omega^{(1)}]{e\cdot fm}.
\eea
It should be noted that in general neutron and proton EDMs can not be related to the meson-nucleon TRIV constants from TRIV potential. However, it is convenient to present expressions for these EDMs, obtained in the chiral limit~\cite{Crewther:1979pi} with the assumption that nucleon EDM are resulted from TRIV potential
\bea
\label{eq:dnp}
d_n = - d_p = \frac{e}{m_N}\frac{g_{\pi}(\bar{g}_\pi^{(0)}-\bar{g}_\pi^{(2)})}{4\pi^2}\ln \frac{m_N}{m_{\pi}}\simeq 0.14 (\bar{g}_\pi^{(0)}-\bar{g}_\pi^{(2)}) ,
\eea
which could be used for some models of CP-violation.

Finally,  one can compare the obtained  expressions for nuclear EDMs with
 TRIV  effects in neutron deuteron elastic scattering related to the $\vs_n\cdot({\vp}\times{\bm I})$ correlation, where  $\vs_n$ is the neutron spin, ${\bm I}$ is the target spin,
and $\vp$ is the neutron momentum, which can be observed in the transmission of polarized neutrons through a target with polarized nuclei.  This correlation leads to the
difference \cite{Stodolsky:1982tp} between the total neutron cross sections  for $\vs_n$
parallel and anti-parallel to ${\vp}\times{\bm I}$
\bea
\Delta\sigma_{\slashed{T}\slashed{P}}=\frac{4\pi}{p}{\rm Im}(f_{+}-f_{-}),
\eea
and to neutron spin rotation angle \cite{Kabir:1982tp} $\phi$  around the axis
${\vp}\times{\bm I}$
\bea
\frac{d\phi_{\slashed{T}\slashed{P}}}{dz}=-\frac{2\pi N}{p}{\rm Re}(f_{+}-f_{-}).
\eea
Here, $f_{+,-}$ are the zero-angle scattering amplitudes for neutrons polarized
parallel and anti-parallel to the ${\vp}\times{\bm I}$ axis, respectively,
 $z$ is the target length, and $N$ is the number of target nuclei per
unit volume. Using results of \cite{Song:2011sw}, one can write
\bea
\label{eq:phiTP}
\frac{1}{N}\frac{d\phi^{\slashed{T}\slashed{P}}}{dz}&=&
(-65 \mbox{ rad}\cdot \mbox{ fm}^2)[\bar{g}_\pi^{(0)}+0.12 \bar{g}_\pi^{(1)}
+0.0072 \bar{g}_\eta^{(0)}+0.0042 \bar{g}_\eta^{(1)} \no & &
-0.0084 \bar{g}_\rho^{(0)}+0.0044 \bar{g}_\rho^{(1)}
-0.0099 \bar{g}_\omega^{(0)}+0.00064 \bar{g}_\omega^{(1)}]
\eea
and
\bea
\label{eq:PTP}
P^{\slashed{T}\slashed{P}}=\frac{\Delta\sigma^{\slashed{T}\slashed{P}}}{2\sigma_{tot}}&=&
\frac{(-0.185 \mbox{ b})}{2\sigma_{tot}}
[\bar{g}_\pi^{(0)}+0.26 \bar{g}_\pi^{(1)}
-0.0012 \bar{g}_\eta^{(0)}+0.0034 \bar{g}_\eta^{(1)} \no & &
-0.0071 \bar{g}_\rho^{(0)}+0.0035 \bar{g}_\rho^{(1)}
+0.0019 \bar{g}_\omega^{(0)}-0.00063 \bar{g}_\omega^{(1)}].
\eea
One can see that both nuclear EDMs and elastic scattering TRIV effects are mostly sensitive to TRIV pion coupling constants. However, while the EDM values are equally sensitive to all isospin parts of the pion coupling constant, the elastic scattering effects are mainly defined  by the isoscalar interactions. This fact clearly demonstrates the complementarity of different TRIV effects in three-nucleon systems. Thus, the relative values of these TRIV parameters may vary for differen models of CP-violation and, therefore, the measurement of a number of TRIV observables  can help to avoid  a possible accidental cancelation of TRIV .


\begin{acknowledgments}
This work was supported by the DOE grant no. DE-FG02-09ER41621.
This work was granted access to the HPC resources of IDRIS
under the allocation 2009-i2009056006
made by GENCI (Grand Equipement National de Calcul Intensif).
We thank the staff members of the IDRIS for their constant help.
\end{acknowledgments}

\bibliography{TViolation}

\end{document}